%% file: main.tex
\documentclass[times,twocolumn]{aastex62}

\usepackage{graphicx}
\usepackage{subfigure}
\usepackage{natbib}
\usepackage{color}
\usepackage{tabularx}
\usepackage{bm}
\usepackage{threeparttable}
\usepackage{hyperref}
\usepackage{amsmath}
\usepackage{multirow}

\newcommand{\thetae}{\theta_{\rm E}}

\newcommand{\pie}{\pi_{\rm E}}

\newcommand{\te}{t_{\rm E}}
\newcommand{\event}{KMT-2023-BLG-1431}

\DeclareGraphicsExtensions{.pdf,.png,.jpg}

\shorttitle{}
\shortauthors{Bell et al.}

\begin{document}
\title{{\large KMT-2023-BLG-1431Lb: A New $q < 10^{-4}$ Microlensing Planet from a Subtle Signature}}

\correspondingauthor{Jiyuan Zhang}
\email{zhangjiyuan2022@gmail.com}

\input{author}

\input{abstract}


\section{Introduction}\label{intro}

\input{intro}

\section{Observations and Data Reduction}\label{obser}

\input{obser}

\section{Light-curve Analysis}\label{model}

\input{model}

\section{Physical Parameters}\label{lens}

\input{lens}

\section{Discussion: The Role of the Follow-up Data}\label{dis}
\input{dis}

\bibliography{Zang.bib}

\end{document}

%% file: author.tex
\author[0000-0003-1946-4852]{Aislyn Bell}
\affiliation{Center for Astrophysics $|$ Harvard \& Smithsonian, 60 Garden St.,Cambridge, MA 02138, USA}
\affiliation{Department of Astrophysical and Planetary Sciences, University of Colorado, Boulder, CO, USA}

\author[0000-0002-1279-0666]{Jiyuan Zhang}
\affiliation{Department of Astronomy, Tsinghua University, Beijing 100084, China}

\author{Youn Kil Jung}
\affiliation{Korea Astronomy and Space Science Institute, Daejon 34055, Republic of Korea}
\affiliation{University of Science and Technology, Korea, (UST), 217 Gajeong-ro Yuseong-gu, Daejeon 34113, Republic of Korea}

\author{Jennifer C. Yee}
\affiliation{Center for Astrophysics $|$ Harvard \& Smithsonian, 60 Garden St.,Cambridge, MA 02138, USA}

\author[0000-0003-0626-8465]{Hongjing Yang}
\affiliation{Department of Astronomy, Tsinghua University, Beijing 100084, China}

\author{Takahiro Sumi}
\affiliation{Department of Earth and Space Science, Graduate School of Science, Osaka University, Toyonaka, Osaka 560-0043, Japan}

\author[0000-0001-5207-5619]{Andrzej Udalski}
\affiliation{Astronomical Observatory, University of Warsaw, Al. Ujazdowskie 4, 00-478 Warszawa, Poland}

\collaboration{(Leading Authors)}

\author{Michael D. Albrow}
\affiliation{University of Canterbury, Department of Physics and Astronomy, Private Bag 4800, Christchurch 8020, New Zealand}

\author{Sun-Ju Chung}
\affiliation{Korea Astronomy and Space Science Institute, Daejon 34055, Republic of Korea}

\author{Andrew Gould}
\affiliation{Max-Planck-Institute for Astronomy, K\"onigstuhl 17, 69117 Heidelberg, Germany}
\affiliation{Department of Astronomy, Ohio State University, 140 W. 18th Ave., Columbus, OH 43210, USA}

\author{Cheongho Han}
\affiliation{Department of Physics, Chungbuk National University, Cheongju 28644, Republic of Korea}

\author{Kyu-Ha Hwang}
\affiliation{Korea Astronomy and Space Science Institute, Daejon 34055, Republic of Korea}

\author{Yoon-Hyun Ryu}
\affiliation{Korea Astronomy and Space Science Institute, Daejon 34055, Republic of Korea}

\author{In-Gu Shin}
\affiliation{Center for Astrophysics $|$ Harvard \& Smithsonian, 60 Garden St.,Cambridge, MA 02138, USA}

\author{Yossi Shvartzvald}
\affiliation{Department of Particle Physics and Astrophysics, Weizmann Institute of Science, Rehovot 76100, Israel}

\author[0000-0001-6000-3463]{Weicheng Zang}
\affiliation{Center for Astrophysics $|$ Harvard \& Smithsonian, 60 Garden St.,Cambridge, MA 02138, USA}

\author{Sang-Mok Cha}
\affiliation{Korea Astronomy and Space Science Institute, Daejon 34055, Republic of Korea}
\affiliation{School of Space Research, Kyung Hee University, Yongin, Kyeonggi 17104, Republic of Korea} 

\author{Dong-Jin Kim}
\affiliation{Korea Astronomy and Space Science Institute, Daejon 34055, Republic of Korea}

\author{Seung-Lee Kim}
\affiliation{Korea Astronomy and Space Science Institute, Daejon 34055, Republic of Korea}

\author{Chung-Uk Lee}
\affiliation{Korea Astronomy and Space Science Institute, Daejon 34055, Republic of Korea}

\author{Dong-Joo Lee}
\affiliation{Korea Astronomy and Space Science Institute, Daejon 34055, Republic of Korea}

\author{Yongseok Lee}
\affiliation{Korea Astronomy and Space Science Institute, Daejon 34055, Republic of Korea}
\affiliation{School of Space Research, Kyung Hee University, Yongin, Kyeonggi 17104, Republic of Korea}

\author{Byeong-Gon Park}
\affiliation{Korea Astronomy and Space Science Institute, Daejon 34055, Republic of Korea}
\affiliation{University of Science and Technology, Korea, (UST), 217 Gajeong-ro Yuseong-gu, Daejeon 34113, Republic of Korea}

\author{Richard W. Pogge}
\affiliation{Department of Astronomy, Ohio State University, 140 W. 18th Ave., Columbus, OH  43210, USA}

\collaboration{(The KMTNet Collaboration)}

\author{Yunyi Tang}
\affiliation{Department of Astronomy, Tsinghua University, Beijing 100084, China}

\author{Jennie McCormick}
\affiliation{Farm Cove Observatory, Centre for Backyard Astrophysics, Pakuranga, Auckland, New Zealand}

\author{Subo Dong}
\affiliation{Department of Astronomy, School of Physics, Peking University, Yiheyuan Rd. 5, Haidian District, Beijing 100871, China}
\affiliation{Kavli Institute for Astronomy and Astrophysics, Peking University, Yiheyuan Rd. 5, Haidian District, Beijing 100871, China}

\author{Zhuokai Liu}
\affiliation{Department of Astronomy, School of Physics, Peking University, Yiheyuan Rd. 5, Haidian District, Beijing 100871, China}
\affiliation{Kavli Institute for Astronomy and Astrophysics, Peking University, Yiheyuan Rd. 5, Haidian District, Beijing 100871, China}

\author{Shude Mao}
\affiliation{Department of Astronomy, Tsinghua University, Beijing 100084, China}

\author{Dan Maoz}
\affiliation{School of Physics and Astronomy, Tel-Aviv University, Tel-Aviv 6997801, Israel}

\author[0000-0003-4027-4711]{Wei Zhu}
\affiliation{Department of Astronomy, Tsinghua University, Beijing 100084, China}

\collaboration{(The MAP \& $\mu$FUN Follow-up Team)}

\author{Fumio Abe}
\affiliation{Institute for Space-Earth Environmental Research, Nagoya University, Nagoya 464-8601, Japan}

\author{Richard Barry}
\affiliation{Code 667, NASA Goddard Space Flight Center, Greenbelt, MD 20771, USA}

\author{David P. Bennett}
\affiliation{Code 667, NASA Goddard Space Flight Center, Greenbelt, MD 20771, USA}
\affiliation{Department of Astronomy, University of Maryland, College Park, MD 20742, USA}

\author{Aparna Bhattacharya}
\affiliation{Code 667, NASA Goddard Space Flight Center, Greenbelt, MD 20771, USA}
\affiliation{Department of Astronomy, University of Maryland, College Park, MD 20742, USA}

\author{Ian A. Bond}
\affiliation{Institute of Natural and Mathematical Sciences, Massey University, Auckland 0745, New Zealand}

\author{Hirosane Fujii}
\affiliation{Department of Earth and Space Science, Graduate School of Science, Osaka University, Toyonaka, Osaka 560-0043, Japan}

\author{Akihiko Fukui}
\affiliation{Department of Earth and Planetary Science, Graduate School of Science, The University of Tokyo, 7-3-1 Hongo, Bunkyo-ku, Tokyo 113-0033, Japan}
\affiliation{Instituto de Astrof\'isica de Canarias, V\'ia L\'actea s/n, E-38205 La Laguna, Tenerife, Spain}

\author{Ryusei Hamada}
\affiliation{Department of Earth and Space Science, Graduate School of Science, Osaka University, Toyonaka, Osaka 560-0043, Japan}

\author{Yuki Hirao}
\affiliation{Department of Earth and Space Science, Graduate School of Science, Osaka University, Toyonaka, Osaka 560-0043, Japan}

\author{Stela Ishitani Silva}
\affiliation{Department of Physics, The Catholic University of America, Washington, DC 20064, USA}
\affiliation{Code 667, NASA Goddard Space Flight Center, Greenbelt, MD 20771, USA}

\author{Yoshitaka Itow}
\affiliation{Institute for Space-Earth Environmental Research, Nagoya University, Nagoya 464-8601, Japan}

\author{Rintaro Kirikawa}
\affiliation{Department of Earth and Space Science, Graduate School of Science, Osaka University, Toyonaka, Osaka 560-0043, Japan}

\author{Iona Kondo}
\affiliation{Department of Earth and Space Science, Graduate School of Science, Osaka University, Toyonaka, Osaka 560-0043, Japan}

\author{Naoki Koshimoto}
\affiliation{Code 667, NASA Goddard Space Flight Center, Greenbelt, MD 20771, USA}
\affiliation{Department of Astronomy, University of Maryland, College Park, MD 20742, USA}

\author{Yutaka Matsubara}
\affiliation{Institute for Space-Earth Environmental Research, Nagoya University, Nagoya 464-8601, Japan}

\author{Sho Matsumoto}
\affiliation{Department of Earth and Space Science, Graduate School of Science, Osaka University, Toyonaka, Osaka 560-0043, Japan}

\author{Shota Miyazaki}
\affiliation{Department of Earth and Space Science, Graduate School of Science, Osaka University, Toyonaka, Osaka 560-0043, Japan}

\author{Yasushi Muraki}
\affiliation{Institute for Space-Earth Environmental Research, Nagoya University, Nagoya 464-8601, Japan}

\author{Arisa Okamura}
\affiliation{Department of Earth and Space Science, Graduate School of Science, Osaka University, Toyonaka, Osaka 560-0043, Japan}

\author{Greg Olmschenk}
\affiliation{Code 667, NASA Goddard Space Flight Center, Greenbelt, MD 20771, USA}

\author{Cl\'ement Ranc}
\affiliation{Sorbonne Universit\'e, CNRS, Institut d'Astrophysique de Paris, IAP, F-75014, Paris, France}

\author{Nicholas J. Rattenbury}
\affiliation{Department of Physics, University of Auckland, Private Bag 92019, Auckland, New Zealand}

\author{Yuki Satoh}
\affiliation{Department of Earth and Space Science, Graduate School of Science, Osaka University, Toyonaka, Osaka 560-0043, Japan}

\author{Daisuke Suzuki}
\affiliation{Department of Earth and Space Science, Graduate School of Science, Osaka University, Toyonaka, Osaka 560-0043, Japan}

\author{Taiga Toda}
\affiliation{Department of Earth and Space Science, Graduate School of Science, Osaka University, Toyonaka, Osaka 560-0043, Japan}

\author{Mio Tomoyoshi}
\affiliation{Department of Earth and Space Science, Graduate School of Science, Osaka University, Toyonaka, Osaka 560-0043, Japan}

\author{Paul J. Tristram}
\affiliation{University of Canterbury Mt.\ John Observatory, P.O. Box 56, Lake Tekapo 8770, New Zealand}

\author{Aikaterini Vandorou}
\affiliation{Code 667, NASA Goddard Space Flight Center, Greenbelt, MD 20771, USA}
\affiliation{Department of Astronomy, University of Maryland, College Park, MD 20742, USA}

\author{Hibiki Yama}
\affiliation{Department of Earth and Space Science, Graduate School of Science, Osaka University, Toyonaka, Osaka 560-0043, Japan}

\author{Kansuke Yamashita}
\affiliation{Department of Earth and Space Science, Graduate School of Science, Osaka University, Toyonaka, Osaka 560-0043, Japan}

\collaboration{(The MOA Collaboration)}

\author[0000-0001-7016-1692]{Przemek Mr\'{o}z}
\affiliation{Astronomical Observatory, University of Warsaw, Al. Ujazdowskie 4, 00-478 Warszawa, Poland}

\author[0000-0002-2335-1730]{Jan Skowron}
\affiliation{Astronomical Observatory, University of Warsaw, Al. Ujazdowskie 4, 00-478 Warszawa, Poland}

\author[0000-0002-9245-6368]{Radoslaw Poleski}
\affiliation{Astronomical Observatory, University of Warsaw, Al. Ujazdowskie 4, 00-478 Warszawa, Poland}

\author[0000-0002-0548-8995]{Micha{\l}~K. Szyma\'{n}ski}
\affiliation{Astronomical Observatory, University of Warsaw, Al. Ujazdowskie 4, 00-478 Warszawa, Poland}

\author[0000-0002-7777-0842]{Igor Soszy\'{n}ski}
\affiliation{Astronomical Observatory, University of Warsaw, Al. Ujazdowskie 4, 00-478 Warszawa, Poland}

\author[0000-0002-2339-5899]{Pawe{\l} Pietrukowicz}
\affiliation{Astronomical Observatory, University of Warsaw, Al. Ujazdowskie 4, 00-478 Warszawa, Poland}

\author[0000-0003-4084-880X]{Szymon Koz{\l}owski}
\affiliation{Astronomical Observatory, University of Warsaw, Al. Ujazdowskie 4, 00-478 Warszawa, Poland}

\author[0000-0001-6364-408X]{Krzysztof Ulaczyk}
\affiliation{Department of Physics, University of Warwick, Gibbet Hill Road, Coventry, CV4~7AL,~UK}

\author[0000-0002-9326-9329]{Krzysztof A. Rybicki}
\affiliation{Astronomical Observatory, University of Warsaw, Al. Ujazdowskie 4, 00-478 Warszawa, Poland}

\author[0000-0002-6212-7221]{Patryk Iwanek}
\affiliation{Astronomical Observatory, University of Warsaw, Al. Ujazdowskie 4, 00-478 Warszawa, Poland}

\author[0000-0002-3051-274X]{Marcin Wrona}
\affiliation{Astronomical Observatory, University of Warsaw, Al. Ujazdowskie 4, 00-478 Warszawa, Poland}

\author[0000-0002-1650-1518]{Mariusz Gromadzki}
\affiliation{Astronomical Observatory, University of Warsaw, Al. Ujazdowskie 4, 00-478 Warszawa, Poland}

\collaboration{(The OGLE Collaboration)}

%% file: abstract.tex
\begin{abstract}

The current studies of microlensing planets are limited by small number statistics. Follow-up observations of high-magnification microlensing events can efficiently form a statistical planetary sample. Since 2020, the Korea Microlensing Telescope Network (KMTNet) and the Las Cumbres Observatory (LCO) global network have been conducting a follow-up program for high-magnification KMTNet events. Here, we report the detection and analysis of a microlensing planetary event, \event, for which the subtle (0.05 magnitude) and short-lived (5 hours) planetary signature was characterized by the follow-up from KMTNet and LCO. A binary-lens single-source (2L1S) analysis reveals a planet/host mass ratio of $q = (0.72 \pm 0.07) \times 10^{-4}$, and the single-lens binary-source (1L2S) model is excluded by $\Delta\chi^2 = 80$. A Bayesian analysis using a Galactic model yields estimates of the host star mass of $M_{\rm host} = 0.57^{+0.33}_{-0.29}~M_\odot$, the planetary mass of $M_{\rm planet} = 13.5_{-6.8}^{+8.1}~M_{\earth}$, and the lens distance of $D_{\rm L} = 6.9_{-1.7}^{+0.8}$~kpc. The projected planet-host separation of $a_\perp = 2.3_{-0.5}^{+0.5}$~au or  $a_\perp = 3.2_{-0.8}^{+0.7}$, subject to the close/wide degeneracy. We also find that without the follow-up data, the survey-only data cannot break the degeneracy of central/resonant caustics and the degeneracy of 2L1S/1L2S models, showing the importance of follow-up observations for current microlensing surveys.

\end{abstract} 

%% file: intro.tex

Gravitational microlensing occurs when a lens star passes in front of a distant source star in an observer's line of sight \citep{einstein1936lens}. The gravitational field from the lens star will alter the path of light rays from the source star, magnifying the source. If a planet is orbiting the lens star near the Einstein radius, it may then perturb the light rays with its gravity. This appears in the data as a deviation from the expected light curve of the star. For typical Galactic microlensing events, the physical Einstein ring radius corresponds to a few AU, so microlensing is most sensitive to planets in these orbits. The two most prolific exoplanet detection methods, the transit and the radial velocity methods, are more sensitive to planets that are close to their host star (e.g., \citealt{pegasib}), so microlensing is complementary to these other detection methods \citep{Mao2012, Gaudi2012}, especially for low mass-ratio ($q \lesssim 10^{-4}$) and wide-orbit planets. 

However, microlensing is a challenging method for detecting exoplanets due to its rare and unpredictable nature. The typical microlensing event rate towards the Galactic bulge is only $\sim 10^{-6}$ \citep{Sumidepth,OGLE4optical}.  Planetary signals within microlensing events are also unpredictable, even rarer, and typically have a duration of one day or less \citep{Shude1991,Andy1992,Bennett1996}. The difficult nature of microlensing has led to uncertainty in the mass-ratio function and multiplicity function. Only one statistical sample \citep{mufun} contains a multi-planet system \citep{OB06109,OB06109_Dave}. 
In addition, the mass-ratio distribution of planets with $\log q < -4$ is still uncertain. A study by \citet{Suzuki2016} contained 22 planet detections, but only two $q < 10^{-4}$ planets. That study found that the number of planets increases as $q$ decreases until $q \sim 1.7\times10^{-4}$, below which the planetary occurrence rate drops.
In order to improve our understand of these planets, it is essential to detect more $q < 10^{-4}$ planets and multi-planet systems in a statistically robust manner that enables population studies. 

An efficient method of detecting microlensing planets is through follow-up observations of high-magnification events. High-magnification events are sensitive to planet detections because planets always produce a ``central" caustic at the position of the lens star, and the source trajectory (by definition for a high-magnification event) passes very close to the lens star \citep{Griest1998}. This also makes them the primary channel for detecting multi-planet systems, because the perturbations from different planets occur near each other in both time and space. In fact, four \citep{Gaudi08_109,OB120026,OB190468,KB211077} out of five unambiguous multi-planet systems were detected in high-magnification events, and a fifth was detected in an event only just barely missing the magnification threshold \citep[$A_{\rm thresh} > 25$, see below;][]{OB181011}. These events additionally have predictable peaks that are usually several magnitudes brighter than the baseline object, making them ideal candidates for follow-up observations. For example, the second microlensing planet, OGLE-2005-BLG-071Lb \citep{OB050071, OB050071D}, was detected by follow-up observations to high-magnification events. In addition, the first measurement of the microlensing planetary frequency was from a follow-up network called the Microlensing Follow Up Network ($\mu$FUN) for high-magnification events \citep{mufun}. 

%

Since the commissioning of KMTNet, microlensing planet detections have been increasingly dominated by detections in the survey data. However, previous work on high-magnification events \citep[e.g., ][]{mufun,MB11293,YeeHung13_311} has suggested that there can be a higher threshold for planet detections in such events because the data characterizing the planet anomalies can overlap with the data that characterizes the underlying event. Hence, even with high-cadence survey data, high-magnification events can benefit from additional monitoring.

Since July 2020, the Microlensing Astronomy Probe (MAP\footnote{\url{http://i.astro.tsinghua.edu.cn/~smao/MAP/}}) collaboration has been using the Las Cumbres Observatory global network (LCO) to systematically conduct follow-up observations of high-magnification microlensing events \citep{LCOGT}. In addition to LCO, this program also uses $\mu$FUN and the Korean Microlensing Telescope Network (KMTNet, \citealt{KMT2016}) to take follow-up observations. The KMTNet AlertFinder system supports this project by releasing new microlensing events every working day and updating the photometry every three hours (\citealt{KMTAF}). This event-alert system, combined with the HighMagFinder system \citep{KB210171}, identifies high-magnification events before they reach the magnification threshold of $A_{\rm thresh} = 25$ for follow-up\footnote{Although early follow-up work used a threshold  $A_{\rm thresh} = 100$, this limit was partially due to limitations in observing resources. Work by \citet{Abe13} and \citet{OB190960} has shown that $A_{\rm thresh} = 25$ is better for capturing the maximum sensitivity of this class of events, although it requires observing more events for a longer duration.}. The data from this follow-up project has been used in the papers of nine planets \citep{KB200414,KB210171,MB20208,KB220440,KB210912,KB211547,KB220371}. Among them, KMT-2020-BLG-0414Lb has the lowest mass ratio ($q=(0.9-1.2)\times10^{-5}$) of the microlensing planets detected thus far. In 2023, we continue our follow-up project and detected another low-$q$ planet, KMT-2023-BLG-1431Lb, which has a mass ratio of $q = (0.72 \pm 0.07) \times 10^{-4}$.

The paper is structured as follows. In Section \ref{obser} we introduce the survey and follow-up observations for this event. In Section \ref{model}, we present the binary-lens single-source (2L1S) and single-lens binary source (1L2S) analysis. In Section \ref{lens}, we conduct a color-magnitude diagram (CMD) analysis and a Bayesian analysis to estimate the
lens physical parameters. Finally, we investigate the results only using the survey data and discuss the implications of this work in Section \ref{dis}. 


%% file: obser.tex
\begin{table*}[ht]
    \renewcommand\arraystretch{1.2}
    \centering
    \caption{Data information with corresponding data reduction method}
    \begin{threeparttable}
    \begin{tabular}{c c l c r c c c c}
    \hline
    Collaboration & Site & Name & Filter & $N_{\rm data}$ & Reduction Method & $(k,e_{\rm{min}})$\tnote{1} \\
    \hline
    
    KMTNet & SSO & KMTA04 & $I$ & 398 & pySIS\tnote{2} & $(1.12, 0.004)$ \\
       
    KMTNet & CTIO & KMTC04 & $I$ & 664 & pySIS & $(1.10, 0.004)$ \\
    
    KMTNet & CTIO & KMTC04 ($V$)\tnote{3} & $V$ & 65 & pySIS & ... \\
    
    KMTNet & SAAO & KMTS04 & $I$ & 377 & pySIS & $(1.04, 0.008)$\\
    
    MOA & Mt. John Observatory & MOA & Red & 570 & \cite{Bond2001} & $(1.45, 0.006)$\\ 

    OGLE & Las Campanas Observatory & OGLE & $I$ & 197 & \cite{Wozniak2000} & $(1.83, 0.003)$  \\
    
    MAP & SSO & LCOA & $I$ & 115 & pySIS & $(1.19, 0.002)$\\
    
    MAP & CTIO & LCOC& $I$ & 109 & pySIS & $(0.77, 0.005)$\\
    
    MAP & SAAO & LCOS & $I$ & 143 & pySIS & $(0.87, 0.004)$\\
    
    $\mu$FUN & Farm Cove Observatory & FCO & unfiltered & 45 & pySIS & $(0.46, 0.000)$ \\
    $\mu$FUN & El Sauce Observatory & CHI-18\tnote{4} & 580--700 nm & 212 & pySIS & ... \\
    \hline
    \end{tabular}
   \label{data}
   \begin{tablenotes}
    \item[1] $(k,e_{\rm{min}})$ are the error renormalization factors as described in \cite{MB11293}.
    \item[2] \cite{pysis,Yang_TLC}
    \item[3] Only used for the color measurement of the source star.
    \item[4] Not included in the analysis due to the too low SNR and no coverage on the anomaly.
    \end{tablenotes}
    \end{threeparttable}
\end{table*}

\subsection{Survey Observations}

On June 27 2023 (${\rm HJD}^{\prime} = 10122.5, {\rm HJD}^{\prime} = {\rm HJD} - 2450000$), \event\ was flagged as a clear microlensing event by the KMTNet AlertFinder system \citep{KMTAF}. The event lies in the KMTNet BLG04 field and is located at equatorial coordinates of $(\alpha, \delta)_{\rm J2000}$ = (18:04:44.05, $-$29:44:38.11) and Galactic coordinates of $(\ell,b) = (-1^{\circ}.40, -4^{\circ}.00)$, with a cadence of $1.0~{\rm hr}^{-1}$ \citep{KMTeventfinder}. \event\ was later found by the Microlensing Observations in Astrophysics (MOA, \citealt{Sako2008}) group as MOA-2023-BLG-291 on July 5 2023 \citep{Bond2001} and by the Optical Gravitational Lensing Experiment (OGLE, \citealt{OGLEIV}) group as OGLE-2023-BLG-0879 on July 7 2023. The cadence for the MOA and the OGLE surveys are $\sim 0.7~{\rm hr}^{-1}$, and 0.5--1.0 ${\rm night}^{-1}$, respectively. 

KMTNet consists of three identical telescopes in the southern hemisphere: the Cerro Tololo Inter-American Observatory (CTIO) in Chile (KMTC), the South African Astronomical Observatory (SAAO) in South Africa (KMTS), and the Siding Spring Observatory (SSO) in Australia (KMTA). The KMTNet telescope is 1.6 m and equipped with $4~{\rm deg}^2$ cameras. The MOA group conducted a microlensing survey using a 1.8 m telescope equipped with a 2.2 ${\rm deg}^2$ FoV camera at the Mt. John University Observatory in New Zealand. The OGLE data were acquired using the 1.3m Warsaw Telescope with a 1.4 ${\rm deg}^2$ FoV camera at the Las Campanas Observatory in Chile. Most KMTNet and OGLE observations were made in the $I$ band due to its high signal-to-noise ratio (SNR) for the extincted Bulge fields. A subset of observations in the $V$ band were taken to measure the source color. The MOA images were mainly taken in the MOA-Red band, which is roughly the sum of the standard Cousins $R$ and $I$ band.

\subsection{Follow-up Observations}
At ${\rm HJD}^{\prime} = 10129.4$, i.e., nine days before the highest magnification, the KMTNet HighMagFinder system found that this event is a candidate high-magnification event. Following the alert, the LCO, KMTNet, and $\mu$FUN groups conducted follow-up observations. For LCO, the high-cadence follow-up observations began at ${\rm HJD}^{\prime} = 10137.4$. From ${\rm HJD}^{\prime} = 10138.2$ to $10139.2$, the KMTNet used ``auto-followup'' to increase the cadence of observations for BLG04 by replacing the BLG41 observations ($\Gamma = 1.5~{\rm hr}^{-1}$ for KMTS and KMTA, and $\Gamma = 2.0~{\rm hr}^{-1}$ for KMTC) with BLG04. The $\mu$FUN group took follow-up observations from a 0.18 m Newtonian telescope at El Sauce Observatory in Chile (CHI-18) and the Farm Cove Observatory (FCO) in New Zealand.

\subsection{Data Reduction}

The data used in the light-curve analysis were reduced by the difference image analysis (DIA, \citealt{Tomaney1996, Alard1998}) pipelines: pySIS \citep{pysis,Yang_TLC} for KMTNet, LCO, and $\mu$FUN; \cite{Bond2001} for MOA; and \cite{Wozniak2000} for OGLE. Ultimately, the CHI-18 data were taken after the anomaly and their SNR was too low to constrain the model, so they were not used in the analysis. The $I$-band magnitude of the data has been calibrated to the standard $I$-band magnitude using the OGLE-III star catalog \citep{OGLEIII}. The errors from the DIA pipelines were re-normalized using the method of \cite{MB11293}, which enables $\chi^2/{\rm dof}$ for each data set to become unity, where ``dof'' is the number of degrees of freedom. Table \ref{data} summarizes the reduction method, the error renormalization factors for each data set.

%% file: model.tex

\begin{figure*}[htb] 
    \centering
    \includegraphics[width=0.85\textwidth]{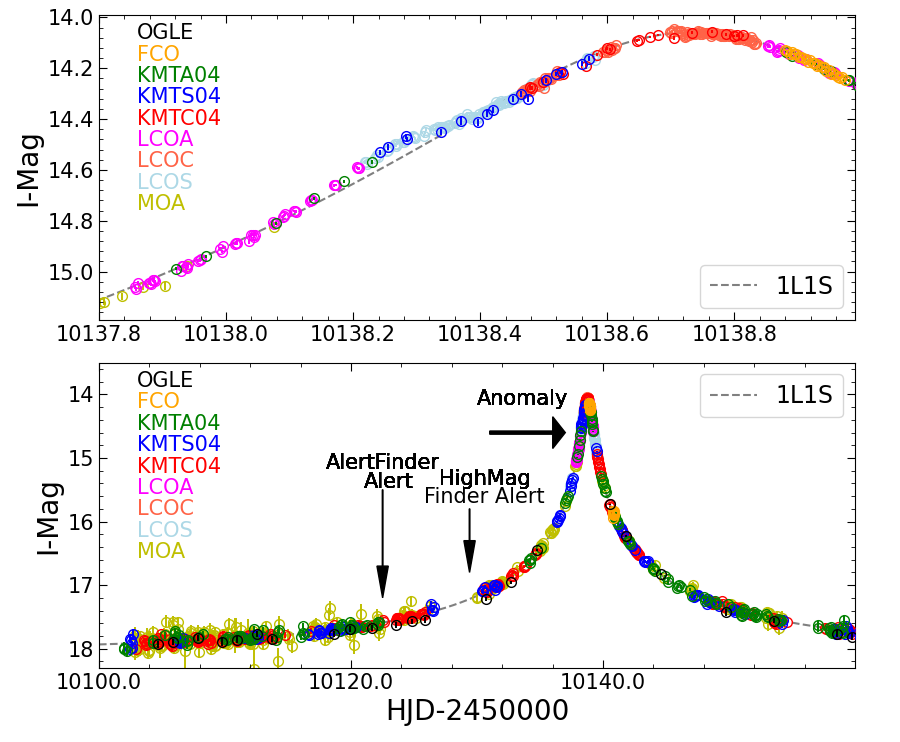}
    \caption{Light curve of \event. {\it Upper:} The event was identified as high-magnification well before the peak, leading to dense observational coverage. {\it Lower:} There is a clear deviation from a PSPL light curve (gray dashed line).}
    \label{fig:lc1}
\end{figure*}

Figure \ref{fig:lc1} displays the observed data together with the best-fit  single-lens single-source (1L1S, \citealt{Paczynski1986}) model. There is a 
0.2 day bump 0.45 day before the peak of the 1L1S model. This anomaly is covered by multiple sites (KMTA04, KMTS04, LOCA, and LCOS) making it very secure. Because such a short-lived bump can be caused by both a binary-lens single-source (2L1S) model and a single-lens binary-source (1L2S) model, we conduct both 2L1S and 1L2S analysis below.

\subsection{Binary-lens Single-source Analysis}

\begin{figure}
    \centering
    \includegraphics[width=0.47\textwidth]{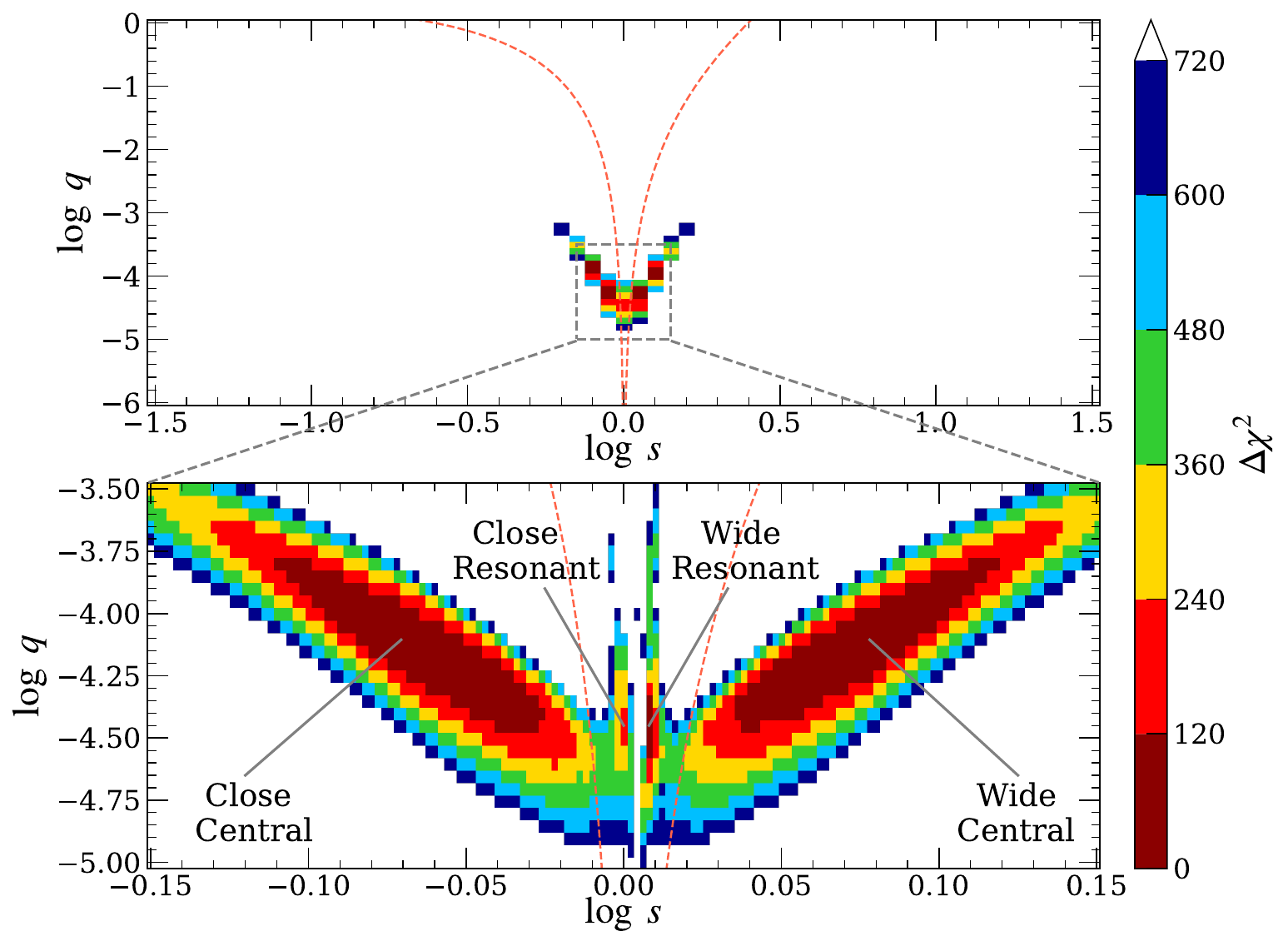}
    \caption{$\chi^2$ surface in the ($\log s, \log q$) plane drawn from the grid search. The upper panel displays the space that is equally divided on a ($61 \times 61$) grid with ranges of $-1.5\leq\log s \leq1.5$ and $-6.0\leq \log q \leq0$, respectively. The lower panel shows the space that is equally divided on a ($151 \times 31$) grid with ranges of $-0.15\leq\log s \leq0.15$ and $-5.0\leq \log q \leq-3.5$, respectively. Grid with $\Delta \chi^2  > 720$ are marked as blank. The labels ``Close Central'', ``Wide Central'', ``Close Resonant'', and ``Wide Resonant'' in the lower panel indicate four local minima. The two red dashed lines represent the boundaries between resonant and non-resonant caustics applying Equation (59) of \citet{Dominik1999}. 
}
\label{fig:grid}
\end{figure}

A static 2L1S model requires seven parameters to calculate the magnification, $A(t)$, at any given time. The first three are ($t_0$, $u_0$, $\te$), i.e., the time at which the source passes closest to the center of lens mass, $t_0$, the impact parameter of this approach normalized by the angular Einstein radius $\thetae$, $u_0$, and the Einstein radius crossing time, 
\begin{equation}\label{eqn:te} 
\te = \frac{\thetae}{\mu_{\rm rel}}; \qquad \thetae = \sqrt{\kappa M_{\rm L} \pi_{\rm rel}},
\end{equation}
where $\kappa \equiv \frac{4G}{c^2\mathrm{au}} \simeq 8.144 \frac{{\rm mas}}{M_{\odot}}$, $M_{\rm L}$ is the lens mass, and $(\pi_{\rm rel}, \mu_{\rm rel})$ are the lens-source relative (parallax, proper motion).
The next three ($q$, $s$, $\alpha$) define the binary geometry: the binary mass ratio, $q$, the projected separation between the binary components normalized to the Einstein radius, $s$, and the angle between the source trajectory and the binary axis, $\alpha$. The last parameter, $\rho$, is the angular source radius $\theta_*$ normalized by the angular Einstein radius, i.e., $\rho = \theta_*/\thetae$. In addition, for each data set $i$, we introduce two flux parameters $f_{{\rm S},i}$ and $f_{{\rm B},i}$, representing the flux of the source star and any blended flux. Then, the observed flux, $f_{i}(t)$, is 
\begin{equation}
    f_{i}(t) = f_{{\rm S},i} A(t) + f_{{\rm B},i},
\end{equation}
where $A(t)$ is calculated by the advanced contour integration code \citep{Bozza2010,Bozza2018} \texttt{VBBinaryLensing}\footnote{\url{http://www.fisica.unisa.it/GravitationAstrophysics/VBBinaryLensing.htm}}. We also consider the brightness profile of the source star by adopting a linear limb-darkening law \citep{An2002,Claret2011}.

To locate the local $\chi^2$ minima, we first conduct a two-step grid search over the parameter plane ($\log s$, $\log q$, $\log \rho$, $\alpha$). In the first step, a sparse grid search consists 61 values evenly distributed in $-1.5\leq \log s\leq 1.5$, 61 values evenly distributed in $-6\leq \log q \leq 0$, nine values evenly distributed in $-4.0\leq \log\rho \leq-1.6$, and 16 values evenly distributed in $0^{\circ}\leq \alpha < 360^{\circ}$. We find the minimum $\chi^2$ by the Markov chain Monte Carlo (MCMC) $\chi^2$ minimization using the \texttt{emcee} ensemble sampler \citep{emcee}. We fix $\log q$, $\log s$, and $\log \rho$ and let the other four parameters ($t_0, u_0, \te, \alpha$) vary. As shown in the upper panel of Figure \ref{fig:grid}, the minima are contained in the region $-0.15 \leq \log s \leq 0.15$ and $-5.0 \leq \log q \leq -3.5$. In the second step, we thus conduct a denser grid search with 151 values equally spaced between $-0.15\leq \log s\leq 0.15$, 31 values equally spaced between $-5.0\leq \log q \leq -3.5$, seven values evenly distributed in $-3.5\leq \log\rho \leq-2.3$, and 16 initial values evenly distributed in $0^{\circ}\leq \alpha < 360^{\circ}$. As shown in the lower panel of Figure \ref{fig:grid}, there are four distinct local minima, of which two have central caustics and two have resonant caustics. This topology follows the topology of  the ``central-resonant'' caustic degeneracy, which was first systematically identified in 2021 KMTNet season \citep{KMT2021_mass1,KMT2021_mass3,KB210171}.  We label the four solutions as ``Close Central'', ``Wide Central'', ``Close Resonant'', and ``Wide Resonant'', respectively. 

\begin{table*}
    \renewcommand\arraystretch{1.25}
    \centering
    \caption{Lensing parameters for \event\ with the survey and follow-up data}
    \begin{tabular}{c|c c c c|c}
    \hline
    \hline
    \multirow{3}{*}{Parameters} &  \multicolumn{4}{c|}{2L1S} & 1L2S \\
    
     &  \multicolumn{2}{c}{Central} & \multicolumn{2}{c|}{Resonant} & \\ 
      & \multicolumn{1}{c}{Close} & \multicolumn{1}{c}{Wide} & \multicolumn{1}{c}{Close} & \multicolumn{1}{c|}{Wide} & \\
    \hline
    $\chi^2$/dof  & $2591.6/2595$ & $2590.9/2595$ & $2793.8/2595$ & $2623.8/2595$ & $2670.2/2595$ \\
    \hline
    $t_{0,1} - 10138$ (${\rm HJD}^{\prime}$) & $0.736 \pm 0.001$ & $0.735 \pm 0.001$ & $0.736 \pm 0.001$ & $0.736 \pm 0.001$ & $0.745 \pm 0.001$ \\
    $t_{0,2} - 10138$ (${\rm HJD}^{\prime}$) & ... & ... & ... & ... & $0.262 \pm 0.002$ \\ 
    $u_{0,1} (10^{-2})$ & $1.23 \pm 0.02$ & $1.23 \pm 0.02$ & $1.24 \pm 0.02$ & $1.25 \pm 0.02$ & $1.21 \pm 0.01$ \\
    $u_{0,2} (10^{-2})$ & ... & ... & ... & ... & $0.05 \pm 0.07$\\
    $\te$ (days) & $30.3 \pm 0.3$ & $30.4 \pm 0.3$ & $30.3 \pm 0.3$ & $30.0 \pm 0.3$ &  $30.9 \pm 0.3$ \\   
    $\rho_1 (10^{-3})$ & $1.95 \pm 0.30$ & $1.86 \pm 0.38$ & $3.13 \pm 0.07$ & $2.77 \pm 0.10$ & ... \\
    $\rho_{2} (10^{-3})$ & ... & ... & ... & ... & $2.87 \pm 0.13$ \\
    $\alpha$ (degree) & $217.65 \pm 0.16$ & $217.69 \pm 0.16$ & $217.81 \pm 0.13$ & $218.00 \pm 0.15$ & ...  \\
    $s$ & $0.864 \pm 0.012$ & $1.184 \pm 0.018$ & $0.999 \pm 0.001$ & $1.018 \pm 0.001$ & ... \\
    $q (10^{-4})$ & $0.719 \pm 0.069$ & $0.729 \pm 0.073$ & $0.344 \pm 0.016$ & $0.335 \pm 0.012$ & ... \\ 
    $\log q$ & $-4.145 \pm 0.041$ & $-4.140 \pm 0.043$ & $-4.464 \pm 0.020$ & $-4.475 \pm 0.016$ & ...  \\
    $q_{f,I} (10^{-3})$ & ... & ... & ... & ... & $6.21 \pm 0.49$ \\
    $I_{\rm S, OGLE}$ & $18.748 \pm 0.014$ & $18.749 \pm 0.014$ & $18.746 \pm 0.014$ & $18.736 \pm 0.014$ & $18.771 \pm 0.013$ \\ 
    \hline
    \hline
    \end{tabular}
    \label{tab:All_2L1S}
\end{table*}

\begin{figure}
    \includegraphics[width=0.47\textwidth]{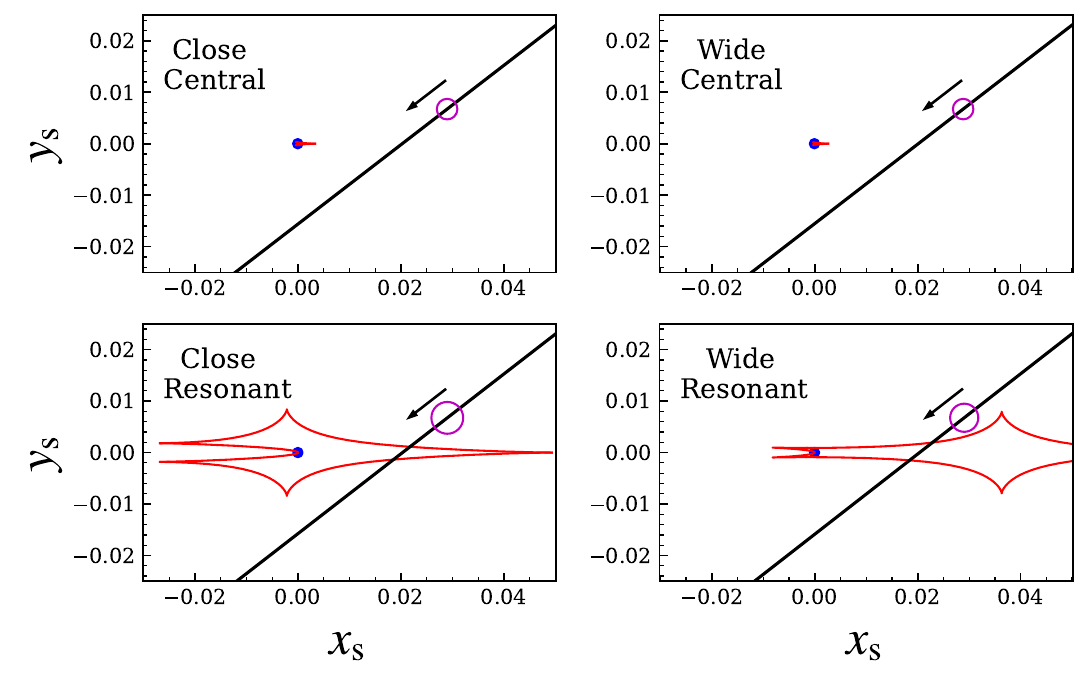}
    \caption{Geometries of the four 2L1S solutions. In each panel, the black line with an arrow represents the source trajectory with respect to the host star that is marked by blue dot, the red lines show the caustic structure, the axes are in units of the Einstein radius $\thetae$, and the magenta circle indicates the source radii.
    }
\label{fig:cau}
\end{figure}

\begin{figure*}
    \centering
    \includegraphics[width=0.85\textwidth]{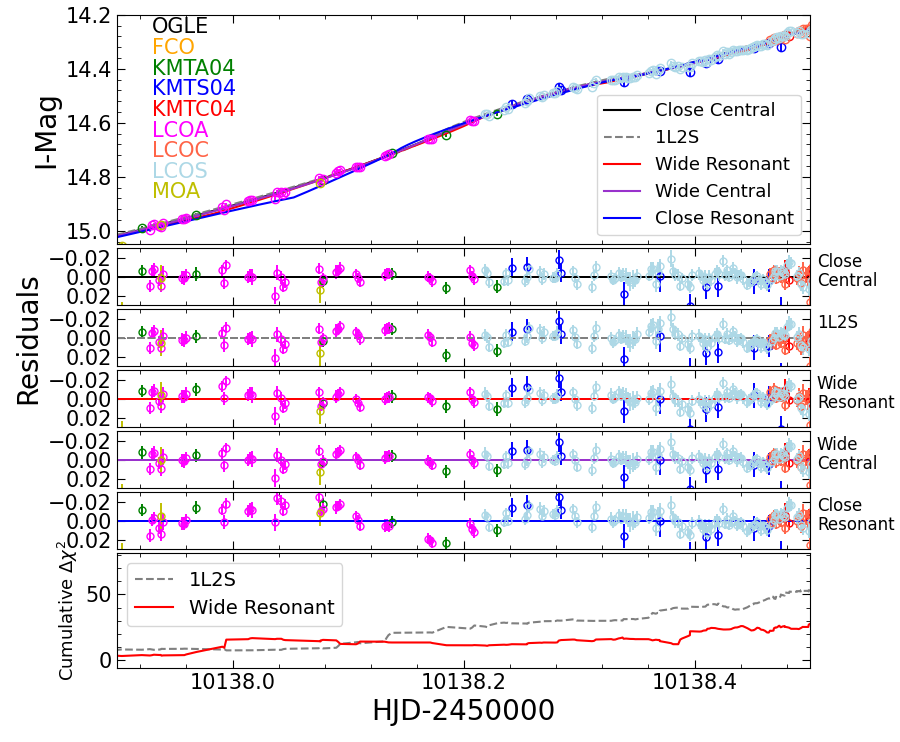}
    \caption{Detailed comparison of the disfavored model fits to the anomaly: ``Close Central", ``Wide Central'', ``Close Resonant" ``Wide Resonant",  and "1L2S". The top panel shows the models plotted with the data, while the middle panels show the residuals to the models. The ``Close Resonant" models shows clear residuals and is ruled out. The deviations in the ``Wide Resonant" and ``1L2S" models are more subtle but (as shown in the bottom panel), amount to $\Delta\chi^2$ differences of $\sim 30$ and $\sim 80$, respectively, over the course of the anomaly. The ``Close Central" model is competitive with the best-fit ``Wide Central" model.}
    \label{fig:lc2}
\end{figure*}

We then investigate the best-fit solutions for each local minimum with all free parameters using the MCMC. Table \ref{tab:All_2L1S} presents the resulting parameters. Figure \ref{fig:cau} displays the caustic geometries and Figure \ref{fig:lc2} shows a close-up of the anomalies together with the model curves. The ``Wide Central'' solution provides the best fit to the observed data, and the ``Close Central'', ``Close Resonant'', and ``Wide Resonant'' solutions are disfavored by $\Delta\chi^2 = 0.7, 203$, and 33, respectively. The ``Close Resonant'' shows significant residuals to the data within the anomaly, so we exclude it. The ``Wide Resonant'' solution does not fit the beginning or the end of anomaly well, and the $\chi^2$ difference is supported consistently by multiple data sets (LCOA, LCOS and KMTA), so we also rule out this solution. Hence, we only consider the two ``Central'' solutions in further analysis. 

We note that, in contrast with other cases of the central-resonant degeneracy, 
(e.g., KMT-2021-BLG-0171, \citealt{KB210171}), for \event, the two ``Resonant" solutions are not degenerate with each other.
In the present case, the ``Close Resonant'' solution is disfavored by $\Delta\chi^2 = 170$ compared to the ``Wide Resonant'' solution. Figure \ref{fig:lc2} shows a close-up of the planetary signal, from which we find that the main difference between the two ``Resonant'' solutions is at the beginning of the anomaly. That is, the ``Close Resonant'' solution shows a slight dip prior to the caustic crossing, while the ``Wide Resonant'' solution exhibits a smooth light curve.

In addition, although the two ``Central'' solutions have no caustic crossings and the separation between the central caustic and the source is about eight times the source radius during the anomaly, $\rho$ is still measured and favored over a point-source model by $\Delta\chi^2 > 15$. This is similar to the central-caustic solution of OGLE-2016-BLG-1195 \citep{OB161195,OB161195_MOA,OB161195_true}, for which $\rho$ was measured (6\% uncertainty) with a separation of 16 times the source radius. 

We also check the microlensing parallax effect \citep{Gould1992, Gould2000} and find a $\chi^2$ improvement of 25. However, the parallax value, $1.8 \pm 0.4$, is of low probability and only the KMTC and KMTS data show the parallax signal. Thus, the suspicious parallax signal is likely due to systematics in the KMTC and KMTS data and we adopt the static models.

\subsection{Single-lens Binary-source Analysis}

\cite{Gaudi1998} suggested that a 1L2S model can also produce a short-lived bump-type anomaly if the second source is much fainter and passes closer to the host star. The total magnification $A_{\lambda}(t)$ for a waveband $\lambda$ is the superposition of the 1L1S magnification of two sources and can be expressed as \citep{MB12486}
\begin{equation}\label{equation2}
    A_{\lambda}(t) = \frac{A_{1}(t)f_{1,\lambda} + A_{2}(t)f_{2,\lambda}}{f_{1,\lambda} + f_{2,\lambda}} = \frac{A_{1}(t) + q_{f,\lambda}A_{2}(t)}{1 + q_{f,\lambda}}, 
\end{equation}
\begin{equation}
    q_{f,\lambda} = \frac{f_{2,\lambda}}{f_{1,\lambda}}, 
\end{equation}
where $f_{j,\lambda}$ and $A_{j}(t)$ ($j = 1, 2$) are the flux at waveband $\lambda$ and magnification of the two sources, respectively. 

We search for the best-fit 1L2S model using MCMC, and the resulting parameters are given in Table \ref{tab:All_2L1S}. The 1L2S model is disfavored by $\Delta\chi^2 \sim 80$ compared to the best-fit 2L1S model. From Figure \ref{fig:lc2}, we find that the $\chi^2$ difference comes mainly from the anomaly, rather than some other source, reinforcing the conclusion that the 1L2S model is a poor fit to the anomaly. In addition, the putative source companion is 5.5 magnitudes fainter than the primary source. According to Section \ref{lens}, the putative secondary source would have an absolute magnitude of $M_{I,2} \sim 9.1$ mag, corresponding to an angular source radius of $\theta_{*,2} \sim 0.2~\mu$as. Then, the lens-source relative proper motion would be $\mu_{\rm rel} = \theta_{*,2}/\rho_2/\te \sim 0.8~{\rm mas\,yr^{-1}}$. Using Equation (9) of \cite{2018_subprime}, which is based on the study of the $\mu_{\rm rel}$ distribution of observed planetary microlensing events \citep{MASADA}, the probability of $\mu_{\rm rel} \leq 0.8~{\rm mas\,yr^{-1}}$ is only 0.018. Hence, based on both the $\Delta\chi^2$ and the low $\mu_{\rm rel}$, we exclude the 1L2S model.

%% file: lens.tex
\subsection{Color-Magnitude Diagram (CMD)}

\begin{figure}[htb] 
    \centering
    \includegraphics[width=0.97\columnwidth]{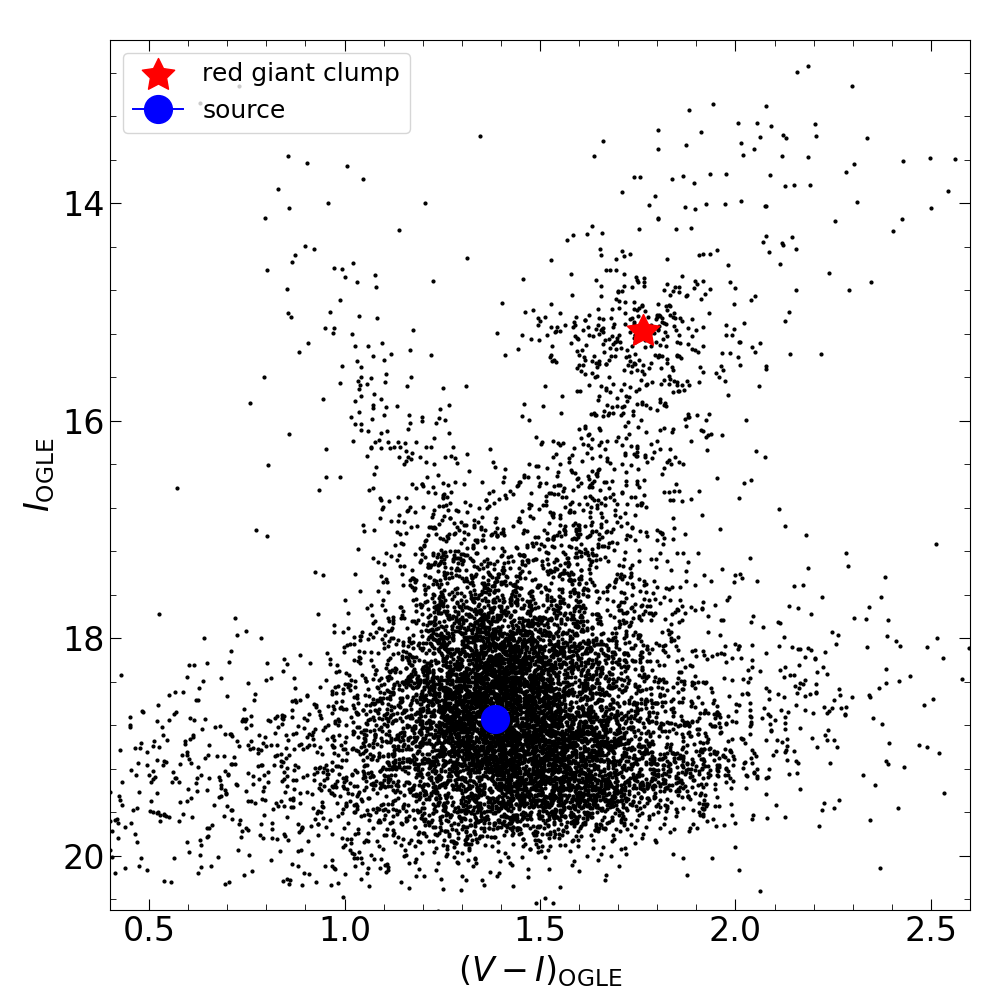}
    \caption{The the OGLE-III CMD for \event, constructed using the field stars within $120^{\prime\prime}$ centered the source star. The red star indicates the centroid of the red giant clump, and the blue dot shows the position of the source star.}
    \label{cmd}
\end{figure}

The 2L1S light-curve analysis yields a measurement of $\rho$, which, combined with the angular source radius $\theta_*$, can be used to calculate the angular Einstein radius: $\thetae = \theta_*/\rho$. We estimate $\theta_*$ by locating the source on a $V - I$ versus $I$ CMD (Figure \ref{cmd}) using the OGLE-III ambient stars \citep{OGLEIII} within $120''$ of the event. The centroid of the red giant clump in this field is measured to be $(V - I, I)_{\rm cl} = (1.76 \pm 0.01, 15.18 \pm 0.02)$. From \cite{Bensby2013} and Table 1 of \cite{Nataf2013}, we estimate the de-reddened color and magnitude of the red giant clump to be $(V - I, I)_{\rm cl,0} = (1.06 \pm 0.03, 14.39 \pm 0.04)$. 

The color and brightness of the source star are measured from the KMTC04 data and converted to the OGLE-III system by matching their respective star catalogs. From the light-curve analysis, the source brightness
is $I=18.75 \pm 0.01$, and the source color, $(V - I)_{\rm S} = 1.76 \pm 0.01$, is derived by regression.
The offsets of these values from the observed red clump leads to the source de-reddened color and magnitude of $(V - I, I)_{\rm S,0} = (0.68 \pm 0.03, 17.96 \pm 0.05)$. According to \cite{Bessell1988}, the source star is probably a G-type dwarf or subgiant. Applying the color/surface-brightness relation for dwarfs and subgiants of \cite{Adams2018}, we obtain the angular source radius of
\begin{equation}
    \theta_* = 0.805 \pm 0.040~\mu{\rm as}. 
\end{equation}
We summarize the CMD parameters and the resulting $\thetae$ and $\mu_{\rm rel}$ for the two ``Central'' solutions in Table \ref{table:source}.

\begin{table}
    \renewcommand{\arraystretch}{1.25}
    \centering
    \caption{CMD parameters and derived $\thetae$ and $\mu_{\rm rel}$ for \event}
    \begin{tabular}{l|r r}
    \hline \hline 
    Red Clump:\\
    $(V - I)_{\rm cl}$ & $1.76 \pm 0.01$ &  \\
    $I_{\rm cl}$ & $15.18 \pm 0.02$ & \\
    $(V - I)_{\rm cl, 0}$ & $1.06 \pm 0.03$ &\\
    $I_{\rm cl, 0}$ & $14.39 \pm 0.04$ & \\
    \hline
    \hline
    & Close Central & Wide Central \\
    \hline
    Source:\\
    $(V - I)_{\rm S}$ & $1.76 \pm 0.01$ & $\gets$ \\
    $I_{\rm S}$ & $18.748 \pm 0.014$ & $18.749 \pm 0.014$ \\ 
    $(V - I)_{\rm S,0}$ & $0.68 \pm 0.03$ & $\gets$ \\
    $I_{\rm S,0}$ & $17.96 \pm 0.05$ & $17.96 \pm 0.05$ \\ 
    $\theta_*$ ($\mu$as) & $0.805 \pm 0.040$ & $0.805 \pm 0.040$ \\
    \hline
    Event:\\
    $\thetae$ (mas) & $0.413 \pm 0.067$  & $0.433 \pm 0.090$ \\ 
    $\mu_{\rm rel}$ (${\rm mas\,yr^{-1}}$) & $4.98 \pm 0.81$ & $5.20 \pm 1.08$ \\ 
    \hline \hline
    \end{tabular}
    \label{table:source}
\end{table}

\subsection{Bayesian Analysis}\label{Baye}

\begin{table*}
    \centering
    \caption{Physical parameters for \event}
    \renewcommand{\arraystretch}{1.4}
    \begin{tabular}{c|cccccc}
    \hline\hline
    \multirow{2}{*}{Solution} & \multicolumn{5}{c}{Physical Properties} &  \\
     & $M_{\rm host}$ ($M_{\odot}$) 
     & $M_{\rm planet}$ ($M_{\oplus}$) 
     & $D_{\rm L}$ (kpc) 
     & $r_{\perp}$ (au) 
     & $\mu_{\rm hel, rel}$ (${\rm mas\,yr^{-1}}$) 
     &  $P_{\rm bulge}$ \\ \hline   
     
    Close Central 
    & $0.57^{+0.32}_{-0.28}$
    & $13.4^{+7.9}_{-6.7}$
    & $6.9^{+0.8}_{-1.6}$
    & $2.3^{+0.5}_{-0.5}$
    & $4.9^{+0.8}_{-0.8}$
    & 65.3\%  \\
    
    Wide Central 
    & $0.57^{+0.33}_{-0.29}$ 
    & $13.6^{+8.3}_{-6.9}$ 
    & $6.8^{+0.8}_{-1.8}$ 
    & $3.2^{+0.7}_{-0.8}$
    & $5.0^{+1.0}_{-1.0}$
    & 62.6\%  \\
    
     \hline\hline
    \end{tabular}
    \begin{tablenotes}
        \centering
      \item[1] $P_{\rm bulge}$ is the probability of a lens in the Galactic bulge.
    \end{tablenotes}
    \label{table:bayes}
\end{table*}

With the angular Einstein radius $\thetae$ and the microlensing parallax $\pie$, the lens mass, $M_{\rm L}$, and the lens distance, $D_{\rm L}$, can be uniquely determined by \citep{Gould1992, Gould2000}
\begin{equation}\label{eq:mass}
    M_{\rm L} = \frac{\thetae}{{\kappa}\pie};\qquad D_{\rm L} = \frac{\mathrm{au}}{\pie\thetae + \pi_{\rm S}},
\end{equation}
where $\pi_{\rm S}$ is the source parallax. For the present case, $\thetae$ is measured but $\pie$ is not constraint, so we estimate the physical parameters of the lens system by a Bayesian analysis based on a Galactic model. 

The Galactic model is the same as used in \cite{KB220440}, in which we adopt initial mass function (IMF) from \cite{Kroupa2001}, with a $1.3M_{\odot}$ and $1.1M_{\odot}$ cutoff for the disk and the bulge lenses, respectively \citep{Zhu2017spitzer}, the stellar number density profile is depicted in \cite{Yang2021_GalacticModel}, and the dynamical distributions of the bulge and disk lenses are described by the \cite{Zhu2017spitzer} and \cite{Yang2021_GalacticModel} model, respectively. 

We generate a sample of $10^7$ simulated events from the Galactic model by conducting a Monte Carlo simulation. For each simulated event $i$ of solution $k$ with parameters $t_{\rm E,i,k}$, $\mu_{{\rm rel},i,k}$, and $\theta_{{\rm E},i,k}$, we weight it by 
\begin{equation}\label{equ:weight}
    w_{i} = \Gamma_{i,k}\times p(t_{\rm E,i,k}) p(\theta_{{\rm E},i,k}),
\end{equation}
where $\Gamma_{i,k} = \theta_{{\rm E},i,k} \times \mu_{{\rm rel},i,k}$ is the microlensing event rate, and $p(t_{\rm E,i,k})$ and $p(\theta_{{\rm E},i,k})$ are the likelihood of $t_{\rm E,i,k}$ and $\theta_{{\rm E},i,k}$, i.e.,
\begin{equation}
\begin{aligned}
& p(t_{\rm E,i,k}) = \frac{{\rm exp}[-(t_{{\rm E},i,k} - t_{{\rm E},k})^2/2\sigma^2_{t_{{\rm E},k}}]}{\sqrt{2\pi}\sigma_{t_{{\rm E},k}}}, \\
& p(\theta_{{\rm E},i,k}) = \frac{{\rm exp}[-(\theta_{{\rm E},i,k} - \theta_{{\rm E},k})^2/2\sigma^2_{\theta_{{\rm E},k}}]}{\sqrt{2\pi}\sigma_{\theta_{{\rm E},k}}},
\end{aligned}
\end{equation}
where $(\sigma_{t_{{\rm E},k}}, \sigma_{\theta_{{\rm E},k}})$ are the standard deviations of $t_{\rm E,k}$ and $\theta_{{\rm E},k}$, respectively. 

Table \ref{table:bayes} presents the resulting posterior distributions of the host mass, $M_{\rm host}$, the planet mass, $M_{\rm planet}$, the lens distance, $D_{\rm L}$, the lens-source relative proper motion in the heliocentric frame, $\mu_{\rm hel, rel}$, the projected planet-host separation, $r_\perp$, derived by $s D_{\rm L} \thetae$, and the probability of a bulge lens, $P_{\rm bluge}$. The values in Table \ref{table:bayes} are the median values of the posterior distributions and the lower and upper limits determined as 16\% and 84\% of the distributions, respectively. It is estimated that the host star is likely an M or K dwarf, in which case the planet mass would be sub-Neptune.


\begin{figure*}[ht]
    \centering
   \includegraphics[width=0.90\textwidth]{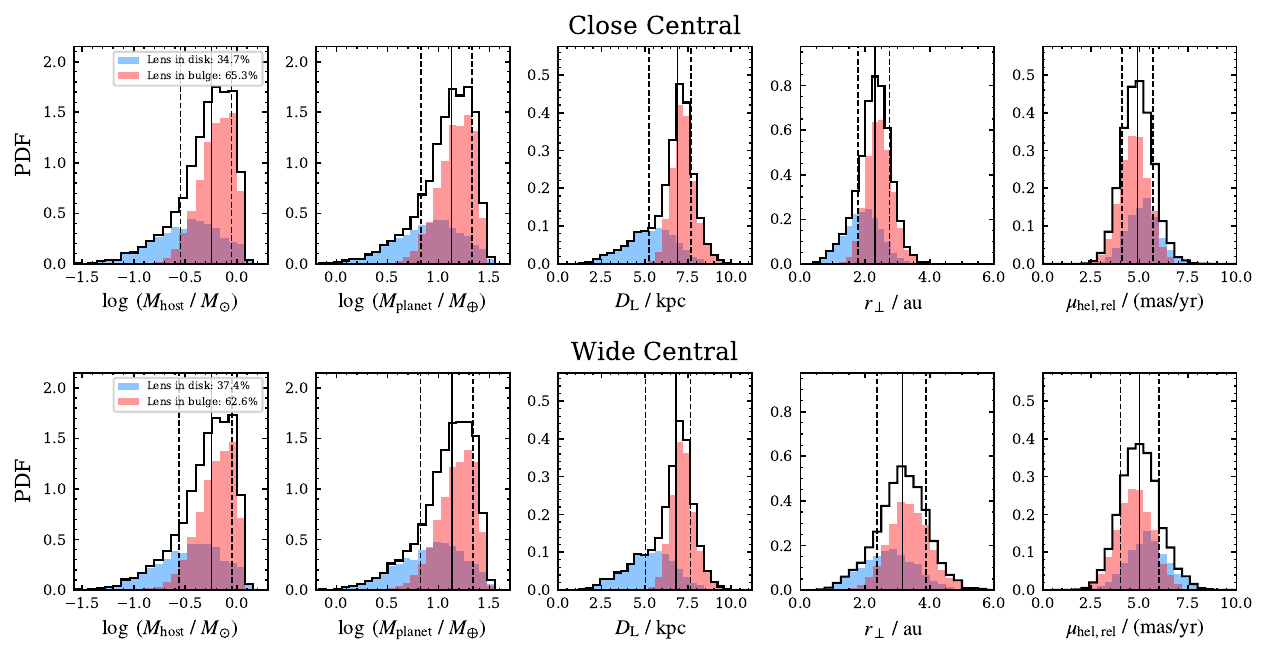}
    \caption{Bayesian posterior distributions for the lens physical parameters of \event. In each panel, the black solid line and the two black dashed lines represent the median value and the 16\% and 84\% percentages of the distribution. Red and blue indicate the distributions for the bulge and disk lenses, respectively.}
\label{fig:bayes}
\end{figure*}

%% file: dis.tex
\begin{figure*}[htb] 
    \centering
    \includegraphics[width=0.85\textwidth]{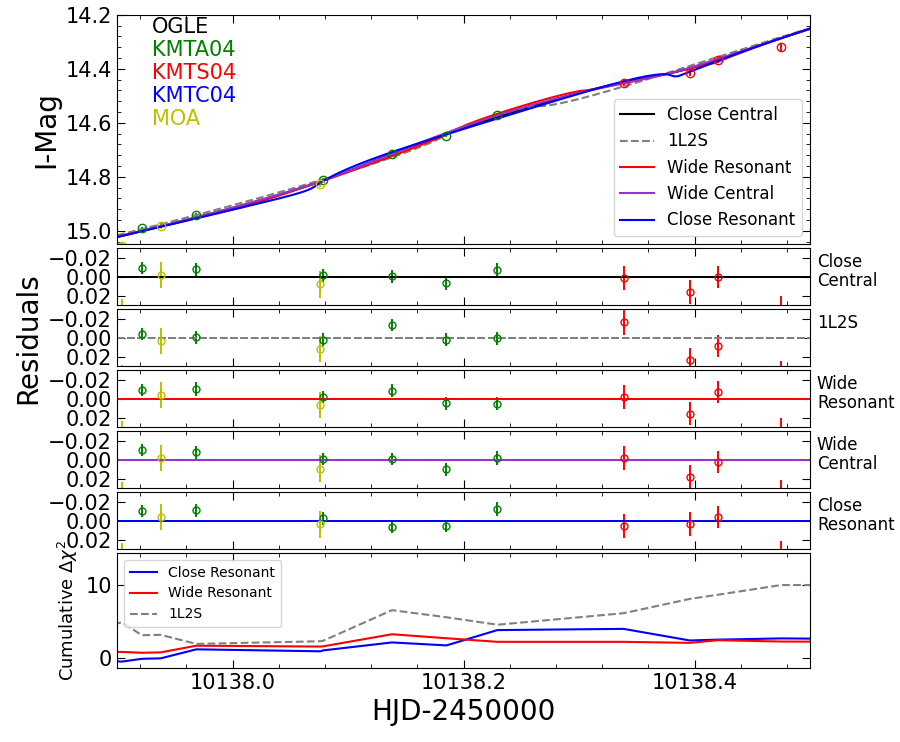}
    \caption{A close-up of the anomaly without the followup data and models fit to only the survey data.}
    \label{fig:lc3}
\end{figure*}

The goal of our follow-up program is to increase the number of planet detections in high-magnification events. This was a case in which the HighMagFinder alerted the event early enough to enable dense observations over the peak, leading to the detection and characterization of a sub-Neptune mass planet. We now consider what would have happened in the absence of our follow-up program.

\event\ lies in KMTNet field BLG04 and so would normally be monitored at a rate of one observation per hour by KMTNet as well as being observed as part of the regular survey operations of OGLE and MOA. To evaluate the ``survey-only" case, we must remove the follow-up observations from LCO and FCO. We must also eliminate the extra KMTNet data that were taken in response to the alert. 

Figure \ref{fig:lc3} shows the light curve in the anomaly region after removing these extra data points. Without the follow-up data, there are only a few points over the bump in the anomaly. In fact, the KMTNet AnomalyFinder algorithm \citep{OB191053,2019_prime}, which operates on the preliminary online pySIS data, on the survey-only KMTNet data cannot find the anomaly. The anomaly fails both the $\Delta\chi^2$ threshold and the requirement that ``at least three successive points $2\sigma$ away from the PSPL model''. So, without the follow-up data, this anomaly would not have been discovered by our automatic algorithm.

On the other hand, high-magnification events are often subject to increased by-eye scrutiny. So assuming that a person could identify the anomaly by eye, we can also ask how well it would be characterized by the survey data alone. 

First, we consider whether or not it would be considered a robust detection, and we find $\Delta\chi^2 = 86.7$ for the best-fit 2L1S model relative to the PSPL model. Although planet detections at this low significance have been published, they tend to be negative perturbations rather than positive ones, because dips in the light curve are considered more robust to correlated noise \citep[cf. OGLE-2018-BLG-0677 with $\Delta\chi^2=46$; ][]{OB180677}. MOA-2010-BLG-311 serves as a counter example: at $\Delta\chi^2 \sim 80$, the anomaly was not considered robust enough to claim a detection \citep{YeeHung13_311}.

Finally, even if the anomaly were considered detected in survey-only data, it would prove difficult to characterize. We repeated the model fits to the survey-only TLC data. The results are given in Table \ref{tab:Survey_2L1S}. This shows that, in the survey-only data, the 1L2S model is only disfavored by $\Delta\chi^2 = 14$, which is marginally excluded, at best. Furthermore, the central/resonant degeneracy cannot be broken, with a maximum $\Delta\chi^2 \sim 3$ between the four solutions. 

In conclusion, our follow-up data play an essential role in both the detection and characterization of this planetary anomaly. This planet, with $q = 0.7\times10^{-4}$, is a perfect example of the class of planets targeted by our systematic follow-up program, and it clearly demonstrates the continued need for such observations, even in the era of wide-field, high-cadence surveys.

\begin{table*}[htb] 
    \renewcommand\arraystretch{1.25}
    \centering
    \caption{Lensing parameters with only the survey data}
    \begin{tabular}{c|c c c c|c}
    \hline
    \hline
    \multirow{3}{*}{Parameters} &  \multicolumn{4}{c|}{2L1S} & 1L2S \\
    
     &  \multicolumn{2}{c}{Central} & \multicolumn{2}{c|}{Resonant} & \\ 
      & \multicolumn{1}{c}{Close} & \multicolumn{1}{c}{Wide} & \multicolumn{1}{c}{Close} & \multicolumn{1}{c|}{Wide} & \\
    \hline
    $\chi^2$/dof  & $2148.0/2156$ & $2149.0/2156$ & $2151.4/2156$ & $2151.3/2156$ & $2162.4/2156$ \\
    \hline
    $t_{0,1} - 10138$ (${\rm HJD}^{\prime}$) & $0.738 \pm 0.001$ & $0.738 \pm 0.001$ & $0.738 \pm 0.001$ & $0.738 \pm 0.001$ & $0.743 \pm 0.002$ \\
    $t_{0,2} - 10138$ (${\rm HJD}^{\prime}$) & ... & ... & ... & ... & $0.227 \pm 0.016$ \\ 
    $u_{0,1} (10^{-2})$ & $1.22 \pm 0.02$ & $1.22 \pm 0.02$ & $1.24 \pm 0.02$ & $1.23 \pm 0.02$ & $1.22 \pm 0.01$ \\
    $u_{0,2} (10^{-2})$ & ... & ... & ... & ... & $0.01 \pm 0.13$ \\
    $\te$ (days) & $30.5 \pm 0.4$ & $30.5 \pm 0.4$ & $30.4 \pm 0.4$ & $30.5 \pm 0.4$ & $30.8 \pm 0.4$ \\   
    $\rho_1 (10^{-3})$ & $<3.5$ & $<3.5$ & $<3.6$ & $<4.0$ & ... \\
    $\rho_2 (10^{-3})$ & ... & ... & ... & ... & $1.89^{+0.83}_{-0.74}$ \\
    $\alpha$ (degree) & $216.47 \pm 0.79$ & $216.47 \pm 0.80$ & $216.54 \pm 0.69$ & $216.91 \pm 0.87$ & ...  \\
    $s$ & $0.886 \pm 0.032$ & $1.161 \pm 0.039$ & $0.986 \pm 0.005$ & $1.018 \pm 0.003$ & ... \\
    $q (10^{-4})$ & $0.431 \pm 0.134$ & $0.439 \pm 0.134$ & $0.178 \pm 0.035$ & $0.227 \pm 0.066$ & ... \\ 
    $\log q$ &  $-4.387 \pm 0.138$ & $-4.378 \pm 0.135$ & $-4.757 \pm 0.088$ & $-4.662 \pm 0.130$ & ... \\
    $q_{f,I} (10^{-3})$ & ... & ... & ... & ... & $3.35 \pm 1.09$ \\
    $I_{\rm S, OGLE}$ & $18.756 \pm 0.014$ & $18.755 \pm 0.014$ & $18.752 \pm 0.014$ & $18.754 \pm 0.014$ & $18.768 \pm 0.015$ \\ 
    \hline
    \hline
    \end{tabular}
    \tablecomments{The upper limit on $\rho$ is $3\sigma$ ($\Delta\chi^2 = 9$).
    }
    \label{tab:Survey_2L1S}
\end{table*}

%
%
%
%
%
%
%

\acknowledgments

W.Zang acknowledges the support from the Harvard-Smithsonian Center for Astrophysics through the CfA Fellowship. J.Z., W.Zang, H.Y., S.M., S.D., Z.L., and W.Zhu acknowledge support by the National  Natural Science Foundation of China (Grant No. 12133005). The SAO REU program is funded in part by the National Science Foundation REU and Department of Defense ASSURE programs under NSF Grant no. AST-2050813, and by the Smithsonian Institution. Work by J.C.Y. and I.-G.S. acknowledge support from N.S.F Grant No. AST-2108414. This research has made use of the KMTNet system operated by the Korea Astronomy and Space Science Institute (KASI) and the data were obtained at three host sites of CTIO in Chile, SAAO in South Africa, and SSO in Australia. Data transfer from the host site to KASI was supported by the Korea Research Environment Open NETwork (KREONET). This research was supported by the Korea Astronomy and Space Science Institute under the R\&D program (Project No. 2023-1-832-03) supervised by the Ministry of Science and ICT. This research uses data obtained through the Telescope Access Program (TAP), which has been funded by the TAP member institutes. This work makes use of observations from the Las Cumbres Observatory global telescope network. The MOA project is supported by JSPS KAKENHI Grant Number JSPS24253004, JSPS26247023, JSPS23340064, JSPS15H00781, JP16H06287, and JP17H02871. Work by RP and JS was supported by Polish National Agency for Academic Exchange grant ``Polish Returns 2019.'' Work by C.H. was supported by the grants of National Research Foundation of Korea (2019R1A2C2085965 and 2020R1A4A2002885). Y.S. acknowledges support from BSF Grant No. 2020740. The authors acknowledge the Tsinghua Astrophysics High-Performance Computing platform at Tsinghua University for providing computational and data storage resources that have contributed to the research results reported within this paper.